\documentclass[letterpaper,12pt]{article}   
\usepackage{osajnl2} 
\usepackage[draft]{hyperref} 
\usepackage{amsmath}
\usepackage{graphicx}
\usepackage{epstopdf}

\begin{document}

\title{Automatic sensitivity-adjustment for a curvature sensor}

\author{Aglae Kellerer, Mark Chun, Christ Ftaclas}
\address{Institute for Astronomy, \\ 
640 N. A'ohoku Place, HI, Hilo 96720, USA\\ 
kellerer@ifa.hawaii.edu}

\begin{abstract} 
There are different techniques  to sense the wavefront phase-distortions due to atmospheric turbulence. Curvature sensors are practical in their sensitivity being adjustable to the prevailing atmospheric conditions. Even at the best sites, the turbulence intensity has been found to vary at times over only a few minutes and regularly over longer periods. Two methods to automatically adjust the sensitivity of a curvature sensor are proposed: First, the defocus distance can be adjusted prior to the adaptive-optics (AO) loop through the acquisition of a long exposure image and can then be kept constant. Secondly, the defocus distance can be changed during the AO loop, based on the voltage values sent to the deformable mirror. We demonstrate that the performance increase -- assessed in terms of the image Strehl-ratio -- can be significant. 
\end{abstract}

\ocis{110.0115, 110.1080} 

\maketitle 

\section{Introduction}

Curvature wavefront sensors\,\cite{roddier1988} have a number of  advantages over the more commonly used Shack-Hartmann sensors\,\cite{SH}. One useful characteristic that has not often been used in practice is their  adjustable sensitivity to phase distortions: While Shack-Hartmann sensors are designed for one fluctuation amplitude taken to be prevalent during the instrument's lifetime, the dynamic range of curvature sensors varies with the distance of the detector from the focal plane (the {\it defocus distance\/}). Accordingly, curvature sensors can be tuned to be optimally sensitive to the wavefront aberrations that prevail during a particular observation. 

In most adaptive-optics applications the amplitude of the phase fluctuations does indeed vary substantially. In astronomy, the seeing -- a measure of the phase-fluctuation amplitudes -- varies on a time scale of typically several minutes, which is  shorter than the integration time of most astronomical observations. The seeing varies even faster and might become prohibitive when an observed object, such as a satellite, moves across the sky faster than astronomical targets.

In 1991, F. Roddier\,\cite{Roddier1} suggested to use a feedback loop to keep the rms tip-tilt signal-error constant. The sensor would then automatically adjust itself to changing seeing conditions and could always operate in its optimum linear range. Existing curvature-sensor based AO systems do not, as yet, implement such a feedback loop. For example on PUEO, the 19 element curvature sensor installed behind the Canda-France-Hawaii telescope, the defocus distance (termed optical gain in the PUEO manual) is not, generally, adjusted to changing atmospheric conditions: ``The optical gain value is (...) by default set to 128 for a point-like object: this value may need to be reduced for an extended object or a double star''\,\cite{PUEOmanual}.
On MACAO\,\cite{MACAO}, the curvature AO systems installed behind the Very Large Telescope,
the defocus distance is adjusted as the reference source is changed. The adjustment depends on the seeing measured by the DIMM-monitor\,\cite{DIMM} and on the apparent size of the reference source (personal communication). 

We propose two practical implementations of defocus feedback-loops: First the defocus distance can be adjusted prior to the AO loop through the acquisition of a long exposure image and can then be kept constant. This solution is similar to the MACAO approach. Secondly, the defocus distance can be changed during the AO loop, based on the voltage values sent to the deformable mirror. We analyze the resultant  performance increase through numerical simulations. 

\section{Optimal defocus distance}\label{sec1}
The output signal of a curvature wavefront sensor is the fractional intensity difference $v=(x-y)/(x+y)$,  where $x$ and $y$ are the photon numbers registered on either side of the focal plane. In closed-loop AO observations $v$ tends to be near 0 and can then be expressed as a function of the wavefront phase, $\phi$\,\cite{NOAO}:
\begin{equation}\label{eq:Roddier}
v(\vec r) = \frac{-\lambda_S f (f-l)}{\pi l} [ \vec{\nabla}(P(\frac{f}{l} \vec r)) \cdot \vec{\nabla}(\phi(\frac{f}{l} \vec r)) + P(\frac{f}{l} \vec r) \cdot \nabla ^2 \phi(\frac{f}{l} \vec r)]
\end{equation}

$\vec r$ is the position vector in the detector plane. 
$f$ is the focal length, $l$ the distance between the detector and the focal plane, and $f/l\,\vec r$ is thus the position vector in the pupil plane. $\lambda_S$ is the sensing wavelength, and $P$ -- the pupil transmission function -- equals 1 inside and 0 outside the pupil.
Eq.\,\ref{eq:Roddier} makes use of the Fresnel approximation (i.e. $D^2/((f-l)\lambda) \geq 1$) and it assumes the size of the diffraction patterns to be negligible compared to the magnitude of the intensity fluctuations in the detector plane. 
This condition,  the geometric-optics approximation, constrains  the defocus distance to  $\lambda_S f^2 << r_0^2(\lambda_S)\, l$. The Fried parameter, $r_0(\lambda_S)$, characterizes the turbulence intensity in dependence on the wavelength, $\lambda_S$. 
In regions where the pupil transmission is uniform, the measured signal is the Laplacian of the wavefront phase. On the pupil edge -- i.a. along the central obstruction and along the telescope spiders -- curvature sensors measure curvature and radial tilt.

In line with Eq.\,\ref{eq:Roddier} we want small defocus distances, $l$, in order to have sufficient sensitivity to small phase fluctuations, but for a given intensity spatial fluctuation size within the aperture, $d = D/\sqrt{N}$ , we want $l$ sufficiently large in order to have $d\,l/ f$ larger than the diffraction pattern $\lambda_S (f-l)/r_0(\lambda_S)$. As Rigaut et al.\,\cite{RigautSPIE} state, to limit the aliasing effect the diffraction should be large to filter out the not measured high-order modes of the turbulence. Accordingly the optimal distance is $d\,l/ f = k_1\,\lambda_S (f-l) / r_0(\lambda_S)$.
The coefficient $k_1$ is sufficiently small to minimize aliasing effects and sufficiently large to avoid non linearities due to diffraction. 
Let $l_0$ be the optimal defocus distance. The defocus length is typically substantially smaller than the telescope focal length, $l<<f$, so that: 

\begin{eqnarray}
l_0 &=& k_1\cdot\frac{ \lambda_S\,f^2\,\sqrt{N}}{D\,r_0(\lambda_S)} \label{eq:lbis}
\end{eqnarray}

the coefficient $k_1$ depends on the exact size of the sub-apertures, and in particular on the size of the central obstruction. But, for a given AO system, $k_1$ is independent of the observing conditions such as seeing and stellar magnitude. 

\section{Defocus adjustment prior to the AO loop}\label{sec:constant}

Here and in the subsequent section, numerical simulations of curvature AO systems will be presented. In this section the seeing is kept constant during each sequence and the point is explored,  whether the optimal defocus distance can be determined from the diameter of a long exposure image recorded prior to the AO loop.

\begin{figure}[htbp]
\begin{center}
\includegraphics[width=0.3\textwidth]{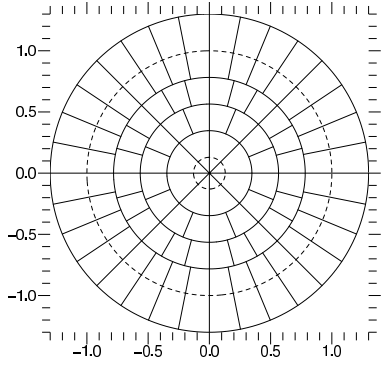}
\caption{Configuration of the simulated sensor: the wavefront is sampled over 80 sub-apertures. Dashed lines: Pupil inner and outer edges.}
\label{fig:apertures}
\end{center}
\end{figure}

The simulated curvature AO-system has $N=80$ sub-apertures, which is an average of existing astronomical curvature systems (PUEO/Canada-France-Hawaii telescope: 19\,\cite{PUEO}, MACAO/Very Large Telescope: 60\,\cite{MACAO}, NICI/Gemini-south telescope: 85\,\cite{NICI}, AO-188/Subaru telescope: 188\,\cite{Subaru}).  The sub-apertures are placed on concentric rings as shown on Fig.\,\ref{fig:apertures}, and cover  equal fractions of the pupil-surface. 

The numerical simulations are based on the code developed by F.\,Rigaut (www.maumae.net/yao/). 
For 22 seeing values between 0.2'' and 1.4\,'' (at $\lambda=0.5\,\mu$m) a matrix of incoming phase values (the {\it phase screen\/}) is generated and  is shifted stepwise in front of the pupil. The distribution of phase values follows the Kolmogorov model. The step-size per iteration is indicated in Table\,\ref{tab:simpar}. Observation sequences are then simulated over 3\,seconds. 

Initially, the optimal defocus length, $l$, and the loop gain, $g$, are determined by repeating the numerical simulations for $g=0.1, 0.2, ... 1$ and $l=0.1, 0.2, ...1.8$\,m in an $f/60$ beam. 
A separate {\it command matrix\/}, which converts a set of $N$ sensor measurements into $N$ voltage values, is computed for each defocus distance. These matrices are computed by first generating $N$ {\it influence functions\/}: $N$ mirror surfaces corresponding to a unit voltage applied to each of the actuators. 
The influence functions are independent of the defocus distance and are generated only once. Note that the influence functions are stored in a $237\times 237$ matrix, while the pupil-diameter equals 220 pixels: all points of the influence functions that lie outside the pupil are set to zero.  
For each defocus distance, {\it Yao\/} computes an {\it interaction matrix\/} through Fresnel diffraction: the $N\times N$ matrix contains the $N$ sensor measurements induced by a unit voltage applied to each of the $N$ actuators. The command matrix is obtained by inverting the interaction matrix. 

\begin{figure}[htbp]
\begin{center}
\includegraphics[width=0.45\textwidth]{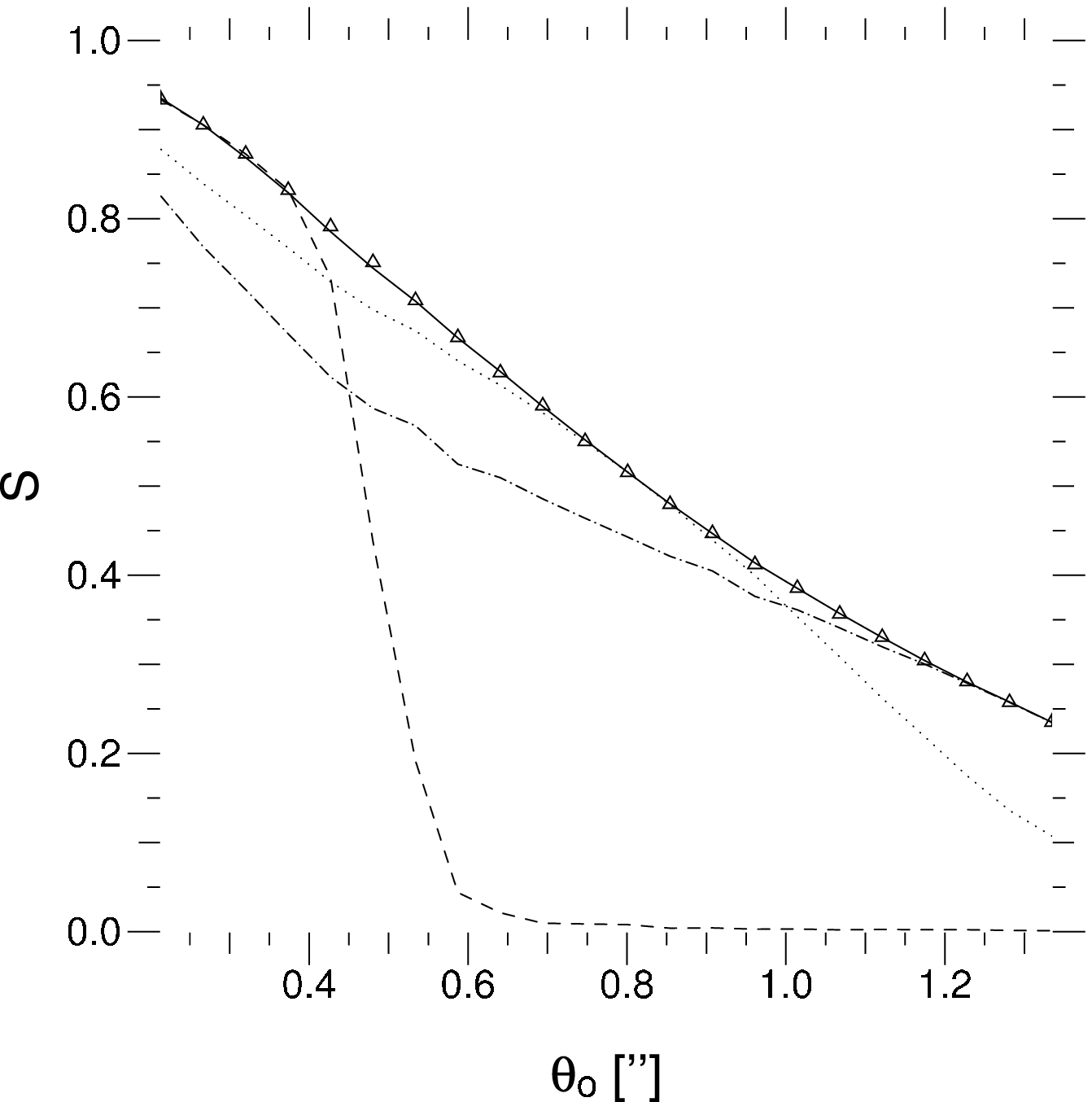}
\includegraphics[width=0.45\textwidth]{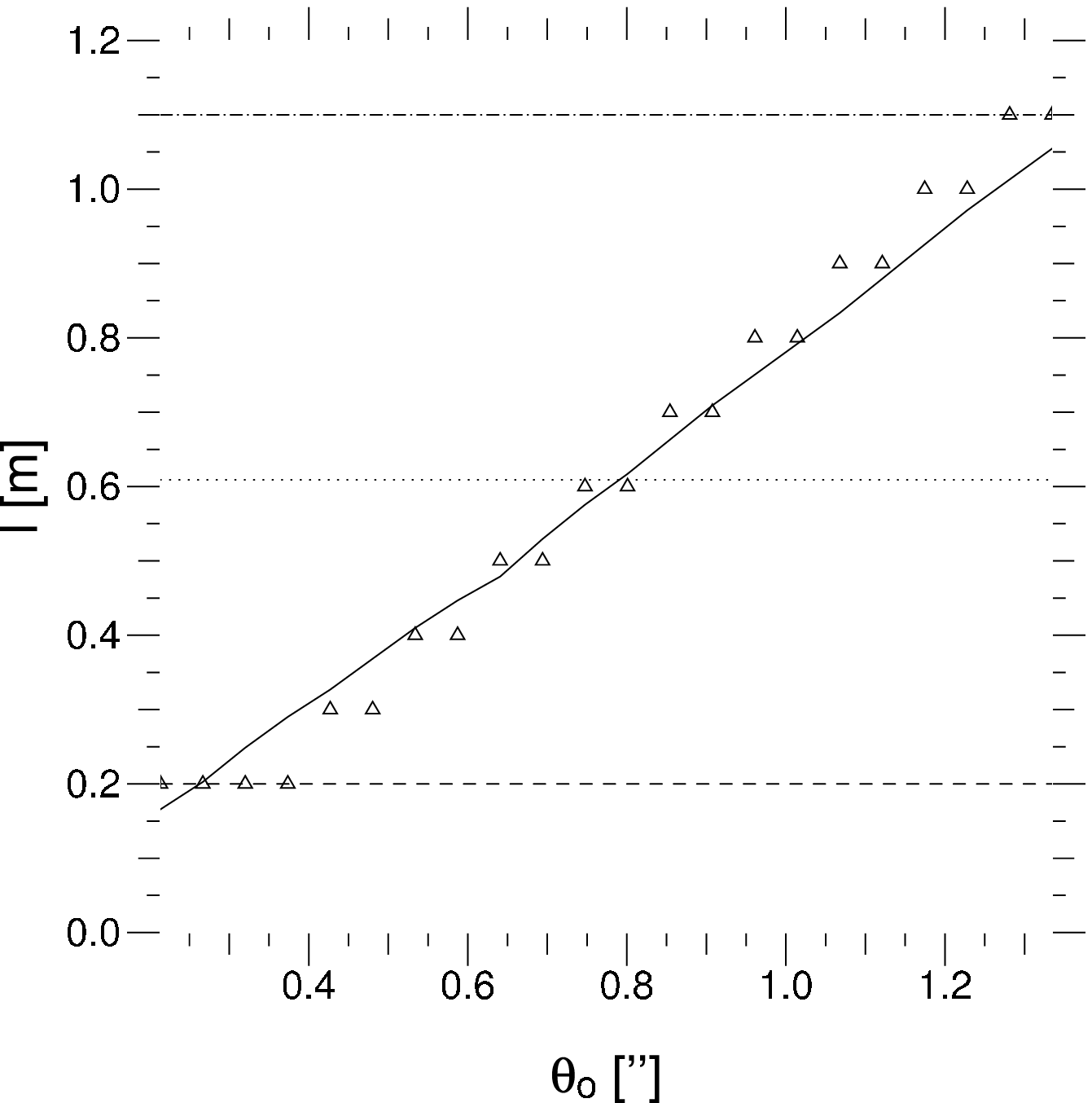}
\caption{Strehl ratio and defocus distance versus seeing for a curvature sensor under various seeing conditions (See left column of Table\,\ref{tab:simpar} for the parameter values).  {\it Triangles\/}: Sequences with different defocus values are simulated and the defocus that yields the highest Strehl is kept.  {\it Dashed line\/} (resp. {\it dotted\/} and {\it dashed-dotted\/}): the minimum (resp. the average and maximum) defocus distance obtained with the former procedure is kept for all turbulence intensities. {\it Solid line:\/} the defocus value is determined from a long-exposure image, acquired prior to the AO loop. }
\label{fig:loop_constant}
\end{center}
\end{figure}

Fig.\,\ref{fig:loop_constant} shows the Strehl ratios attained at $1.65\,\mu$m in dependence on the seeing value at $0.5\,\mu$m and compares them to the ratios achieved when $l$ is kept constant while the seeing changes. In the latter case the loop gain is still optimized for each seeing value.
As shown in Fig.\,\ref{fig:loop_constant},  the correction performance will be especially low if the curvature sensor is tuned for excellent atmospheric conditions (i.e. the defocus distance is small).
The reverse effect -- a performance drop due to an overly pessimistic sensor adjustment -- also exists, but is less fatal: for excellent seeing conditions, $\theta_0=0.2''$, the Strehl ratio decreases from 0.94 ($l$=0.2\,m) to 0.87 ($l$=0.6\,m) and 0.83 ($l$=1.1\,m).

Since it is  impractical to determine the best defocus distance, $l_0$, by running a series of AO loops during actual observations, we need to explore whether -- from the diameter of a long-exposure image acquired prior to the AO loop -- $l_0$ can be directly obtained through the formula: 

\begin{eqnarray}\label{eq:l2}
l_0 &=& k_2\cdot \left( \frac{\lambda_I}{\lambda_S} \right)^{1/5}\cdot\frac{f^2 \sqrt{N}}{D} \cdot  \theta_{50} (\lambda_I)
\end{eqnarray}
$\lambda_S$ and $\lambda_I$ are the sensing and imaging wavelengths. 
$\theta_{50}(\lambda_I)\sim \lambda_I / r_0(\lambda_I)$ [rad] is the diameter of the circle that contains 50\% of the intensity on a long-exposure image at wavelength $\lambda_I$. 
The coefficient $k_2$ depends on instrumental parameters such as the central pupil obstruction and it needs to be determined through simulations or instrumental calibrations.
$D$ is the pupil diameter, $f$ the telescope focal length, and $N$ the number of sub-apertures. Table\,\ref{tab:simpar} gives their numerical values. 

The coefficient $k_2$ is determined via Eq.\,\ref{eq:l2}:  22 values of $l_0$ and $\theta_{50}$ are obtained for seeing values between 0.2'' and 1.4'', at $\lambda=0.5\,\mu$m. A linear fit yields:  
$k_2= 0.68 \pm 0.07$.
As seen on Fig.\,\ref{fig:loop_constant}, the performance of the AO loop is then the same -- whether the defocus is determined prior to the AO sequence from the diameter of a long-exposure image (solid line), or whether many different defocus values are tested (triangles). 
It is thus possible to optimize the system's performance by linking the defocus distance to the diameter of a previously recorded, long exposure image. Another possibility is to adjust the defocus distance in dependence on the seeing measured by a turbulence monitor. On MACAO\,\cite{MACAO} the defocus is adjusted in terms of the seeing as measured by the DIMM\,\cite{DIMM}. 
The disadvantage of this approach is, that the seeing predicted by the DIMM monitor is often much worse than the seeing sensed by the Very Large Telescopes\,\cite{Messenger}. This seems to be explained by  a highly turbulent layer close to the ground, which is seen by the DIMM (installed on a 6\,m high tower), while it does not affect the telescopes (primary mirrors at 11\,m above the ground, top of enclosures at 30\,m).

If the magnitude of the wavefront aberrations changes significantly over the course of the observation, the defocus length needs to be adjusted during the AO loop. In the following section a suitable procedure is suggested and is tested via numerical simulations.

\section{Adjustment during the AO loop}\label{sec:var}

\begin{figure}[htbp]
\begin{center}
\includegraphics[width=0.45\textwidth]{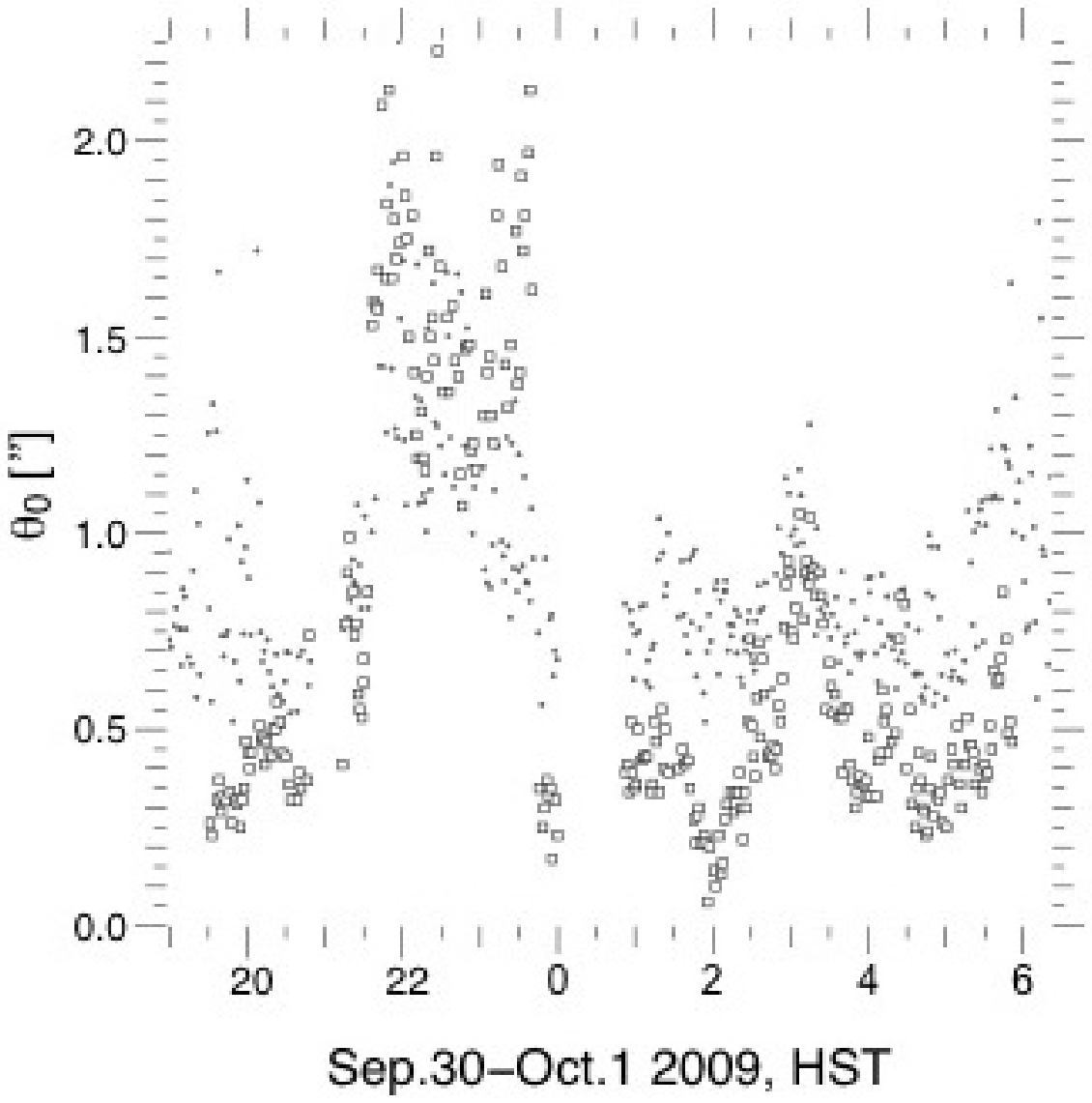}
\includegraphics[width=0.45\textwidth]{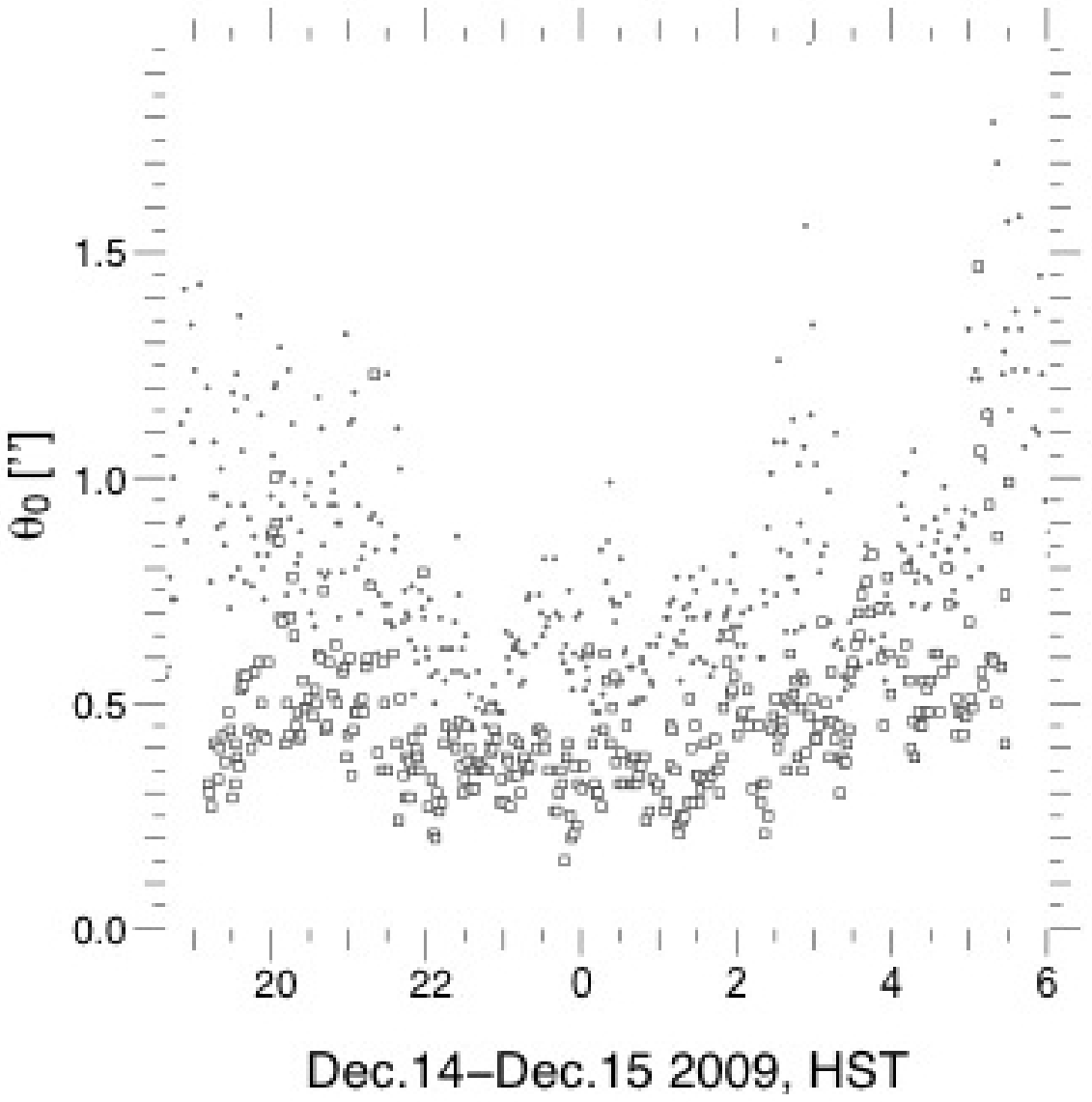}
\caption{Seeing at 0.5\,$\mu$m, recorded by the DIMM (dots) and MASS (squares) on the Mauna Kea observatory. The monitors are installed next to the Canada-France-Hawaii telescope which accommodates the curvature AO system PUEO. DIMM and MASS acquire a measurement about every two minutes. MASS is insensitive to the lowest turbulent layers and generally measures smaller seeing values than DIMM. Credit: http://mkwc.ifa.hawaii.edu/}
\label{fig:seeing}
\end{center}
\end{figure}

Fig.\,\ref{fig:seeing} shows seeing values recorded during two nights by the DIMM\,\cite{DIMM} and MASS\,\cite{MASS} instruments at the Mauna Kea observatory.
Such significant and rapid seeing variations have also been reported at the Paranal observatory by Rigaut and Sarazin\,\cite{RigautSarazin}. The significant and rapid seeing variations suggest that even observations with integration times of just several minutes would benefit from a defocus feedback loop.

\subsection{Method}
In the previous section, the defocus distance has been assumed to be determined prior to the AO loop from the diameter of a long-exposure image. 
To adjust it during the observation, a constantly updated estimate of the seeing is required. 
This estimate is here taken to be obtained from the voltage values sent to the deformable mirror: 

The phase of the wave-front is continuously redetermined from the voltages and the influence functions of the actuators: $\phi = 2\,I\,V$. The voltage vector $V$ has $N$ elements, $N$ being the number of actuators. The matrix $I$ [rad/V] contains the $N$ influence functions. Each influence function is sampled over $M\times M$ points, and $I$ is therefore of size: $M\times M \times N$.  $\phi$ [rad] is the wavefront phase at  wavelength $\lambda_S$ and is likewise sampled over $M\times M$ points.

The phase variance, averaged over a certain number of loop cycles, $K$, is then used to infer the Fried parameter:

\begin{eqnarray}
\sigma_\phi^2 = \frac{1}{K}\cdot \frac{1}{M^2}\,\sum_{k=1}^K \sum_{m,n=1}^M \phi_k(m,n)^2 = \kappa \cdot \left(\frac{D}{r_0(\lambda_S)}\right)^{5/3} 
\end{eqnarray}

$\sigma_\phi^2$ is the average phase variance over $K$ cycles at  wavelength $\lambda_S$. 
$\phi_k$ is the wavefront phase at  wavelength $\lambda_S$ for cycle number $k$. 
The exact value of $\kappa$ need not be assessed, since the defocus distance is related to $D/r_0(\lambda_S)$ through another coefficient (see Eq.\,\ref{eq:lbis}). 
So that:
\begin{eqnarray}
l_0 &=& k_3\cdot \lambda_S \; \left(\frac{f}{D}\right)^2 \; \sqrt{N} \; \sigma_\phi^{6/5}   \label{eq:th}
\end{eqnarray}

$k_3$ is then determined from simulations or instrumental calibrations.  
The actual defocus distance is set to the closest value for which a command matrix has been computed. 

\subsection{Numerical simulations}

At the outset, 30 command matrices are obtained for defocus values in the [0.1, 3.0]\,m range, with a 0.1\,m increment. The computation of the command matrices follows the principle described in Sec.\,\ref{sec:constant}. 
An initial defocus value is specified in terms of Eq.\,\ref{eq:l2} from a long exposure image acquired with the initial seeing value. 

The coefficient $k_3$ is obtained via Eq.\,\ref{eq:th}: Twenty-two AO sequences are simulated for seeing values  between 0.2'' and 1.4'', at $\lambda=0.5\,\mu$m. Each sequence contains 500 iterations. 
The optimal defocus distances corresponding to those seeing values have already been determined in Section\,\ref{sec:constant}. 
 Starting at the 300$^{\rm th}$ iteration, the average phase variance over the last 100 iterations -- i.e. during the last 0.5\,s -- is computed. The mean and standard deviation of the 200 realizations is determined, and the coefficient, $k_3$, is obtained through a linear fit:   $k_3= 1.2\pm 0.1$.

During the AO loop the phase screen is shifted stepwise in front of the pupil. The size of the matrix and the step-size per iteration are indicated in Table\,\ref{tab:simpar}, and allow for a maximum exposure-time of 70\,seconds. Every 5\,seconds the values of the phase screen are scaled to a new seeing value. 
In order to adjust the defocus distance during the AO loop, the wavefront phase variance is computed after each iteration from the eighty voltage-values and their associated influence functions. 

\begin{figure}[htbp]
\begin{center}
\includegraphics[width=0.45\textwidth]{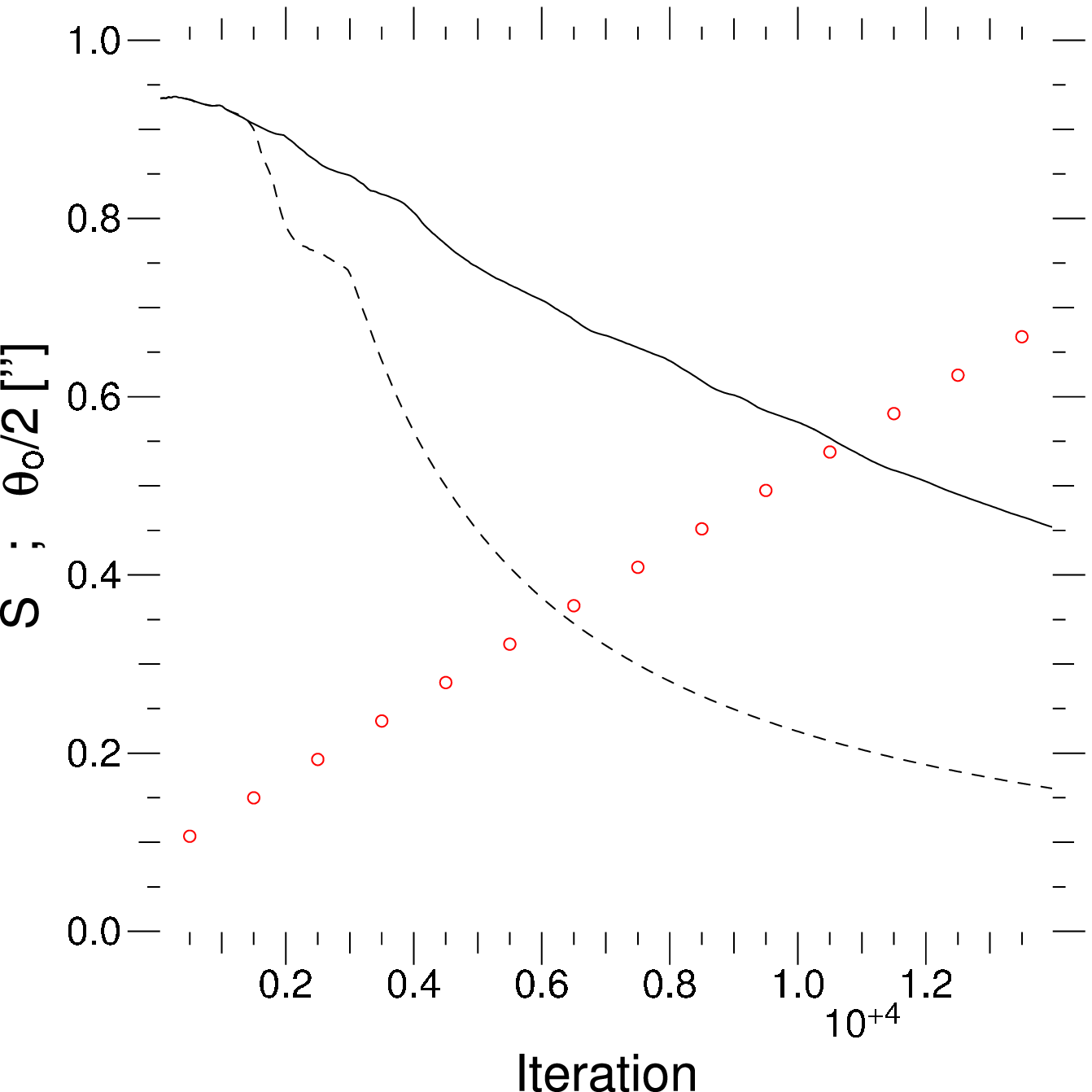}
\includegraphics[width=0.45\textwidth]{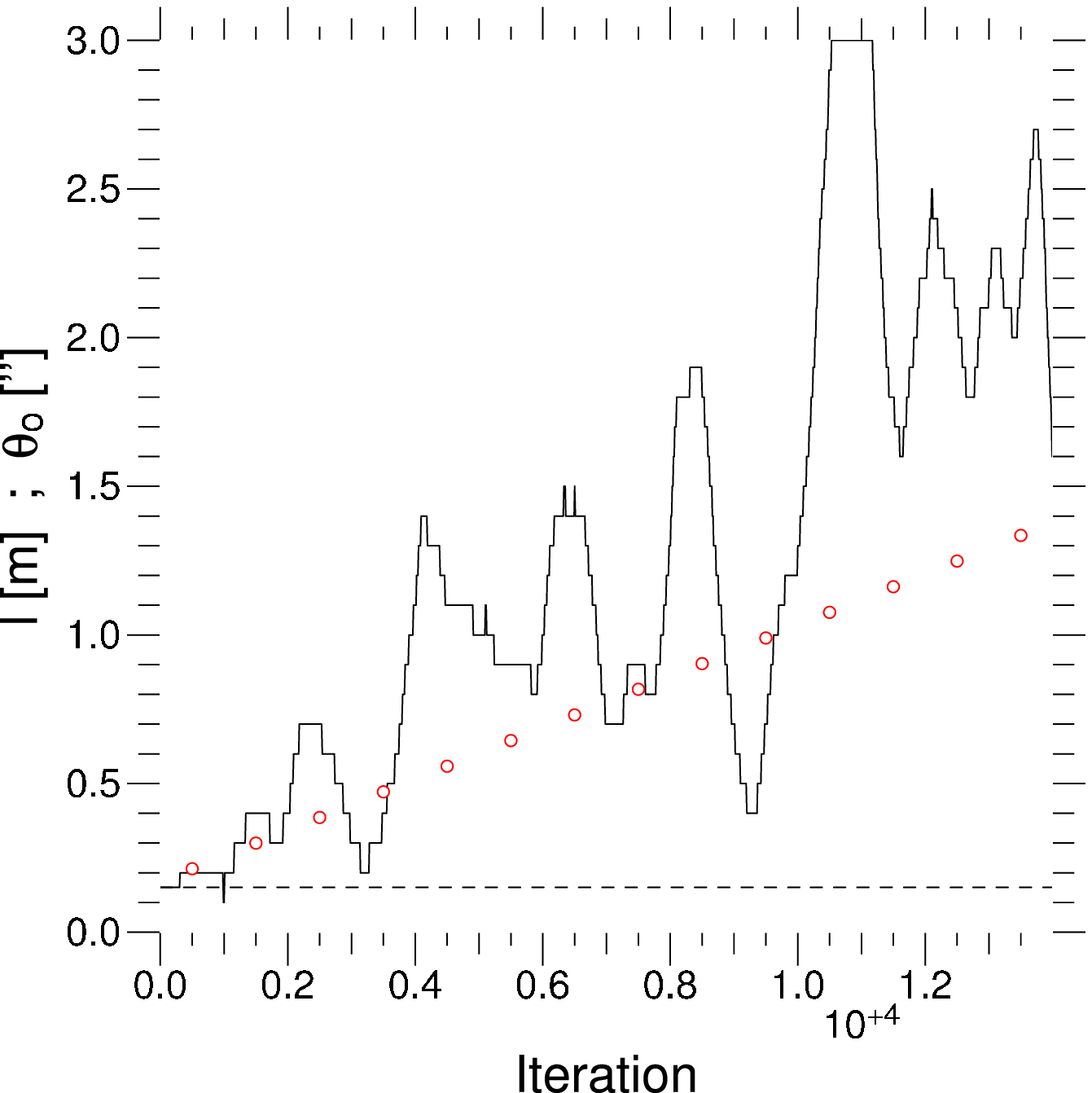}
\caption{
A 70-seconds AO sequence, during which the seeing, $\theta_0$, is incremented by $0.05''$ every 1000$^{\rm th}$ iteration.  \underline{Left panel\/}: {\it Full line\/}: Average Strehl when the defocus is adjusted during the sequence,  {\it dotted lines\/}: the defocus is determined prior to the AO loop when $\theta_0=0.2''$. {\it Circles\/}: $\theta_0/2$. \underline{Right panel\/}: Defocus values. {\it Circles\/}: $\theta_0$. (Parameter values: see right column of Table\,\ref{tab:simpar})}
\label{fig:increasing}
\end{center}
\end{figure}

\begin{figure}[htbp]
\begin{center}
\includegraphics[width=0.45\textwidth]{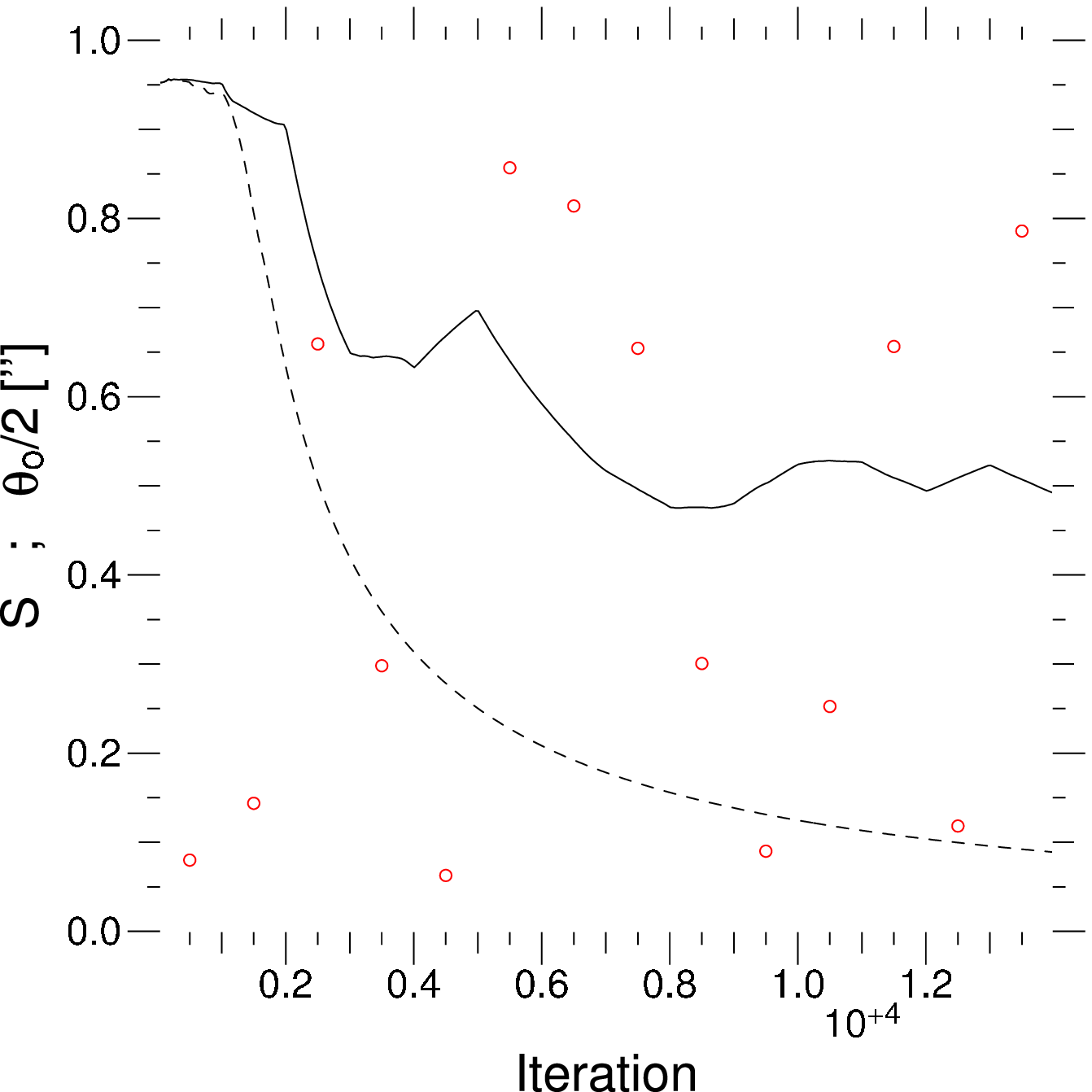}
\includegraphics[width=0.45\textwidth]{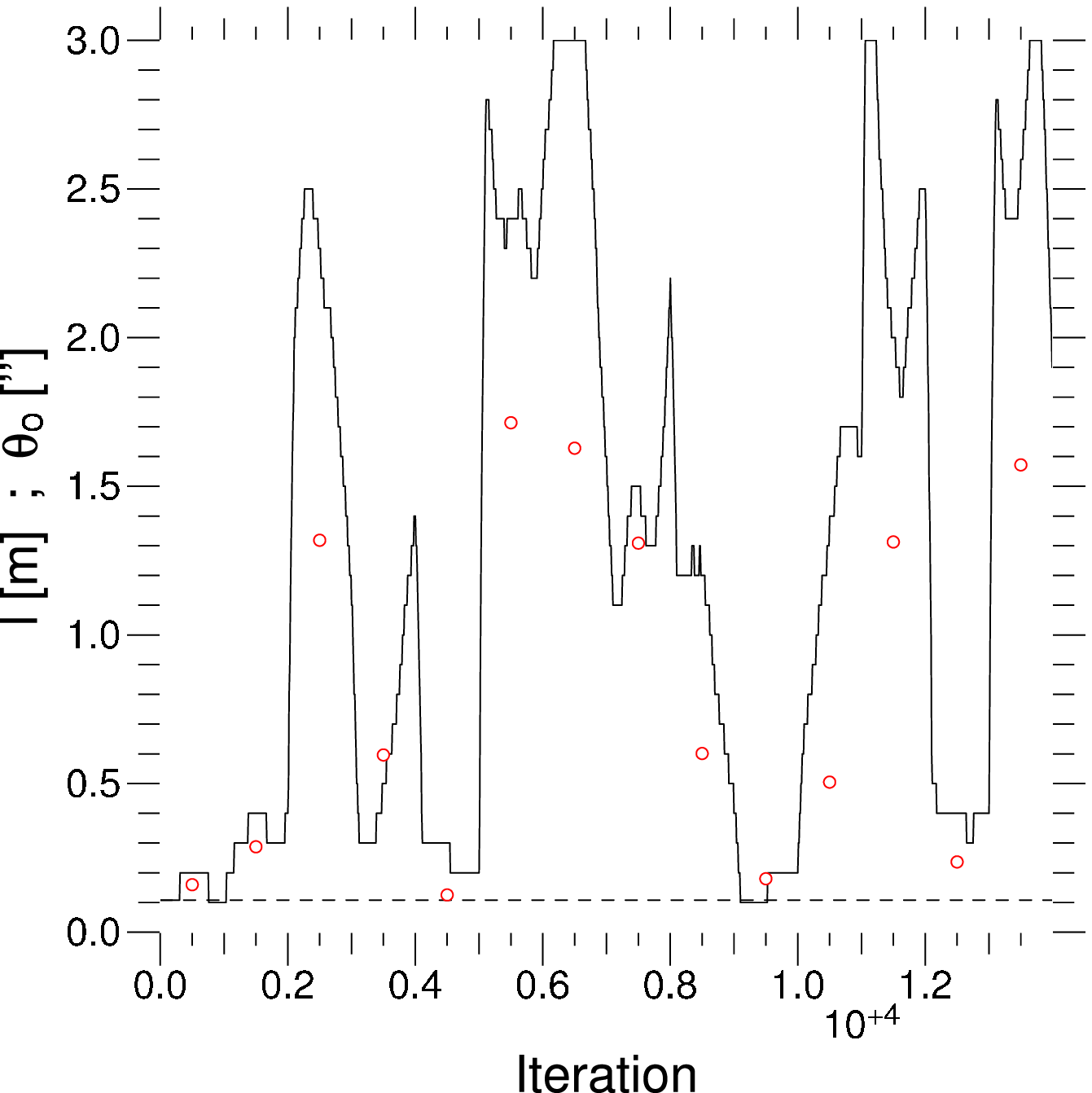}
\caption{
A 70-seconds AO sequence, during which the seeing, $\theta_0$, is modified every 1000$^{\rm th}$ iteration.  \underline{Left panel\/}: {\it Full line\/}: Average Strehl when the defocus is adjusted during the sequence,  {\it dotted lines\/}: the defocus is determined prior to the AO loop when $\theta_0=0.15''$. {\it Circles\/}: $\theta_0/2$. \underline{Right panel\/}: Defocus values. {\it Circles\/}: $\theta_0$. (Parameter values: see right column of Table\,\ref{tab:simpar})}
\label{fig:vary}
\end{center}
\end{figure}

Figures\,\ref{fig:increasing} and \ref{fig:vary} compare the performance of the AO system for two approaches: 
\begin{enumerate}
\item The defocus distance is kept constant. This is the approach presented in Section\,\ref{sec:constant}, but in the present treatment the seeing varies during the observing sequence.

\item Starting from the 300$^{\rm th}$ iteration, the defocus distance is updated every fourth iteration: the average phase variance over the last 100 iterations -- i.e. during the last 0.5\,s -- is used to determine a new defocus distance through Eq.\,\ref{eq:th}. The defocus is set to the closest value for which an interaction matrix has been computed, i.e. it is set to 0.1, 0.2,... 2.9 or 3.0\,m.
\end{enumerate}

In both cases, the AO loop gain is adjusted for the average seeing over the sequence and is then kept constant. 
In Fig.\,\ref{fig:increasing} the seeing increases continuously from 0.2'' to 1.4''. The long exposure image is acquired when $\theta_0=0.2''$ and the initial defocus length is thus small (0.2\,m). Under this condition the Strehl value breaks down when the seeing increases while the defocus distance is kept constant at its low value. Adapting the defocus to the changing turbulence conditions, makes the final Strehl value increase to 0.45. The right panel of Fig.\,\ref{fig:increasing} shows that the defocus increases when the seeing, and hence the phase distortions, increase. However, as the deformable mirror corrects the wavefront, the fluctuation sizes decrease and the defocus distance can again be reduced. This possibility of decreasing the defocus distance as the AO correction takes effect has been discussed F. Roddier\,\cite{RoddierOpCom}.

In Fig.\,\ref{fig:vary}, the seeing varies randomly between 0.1'' and 1.8''. When the defocus is kept constant at its initial low value, the sensor measurements become dominated by diffraction as the seeing increases and erratic voltage commands distort the mirror surface. As the seeing decreases again, the distorted mirror surface still cause large diffraction patterns and the turbulence induced phase-distortions can not be measured: the AO loop has become unstable and the Strehl continuously decreases. When the defocus is adjusted, the average Strehl converges to the Strehl value (0.5) that roughly corresponds  to the mean seeing (0.8''): indeed Fig.\,\ref{fig:loop_constant} shows that the Strehl equals 0.5 when the seeing equals 0.8'' during the whole AO correction. 

\begin{figure}[htbp]
\begin{center}
\includegraphics[width=0.45\textwidth]{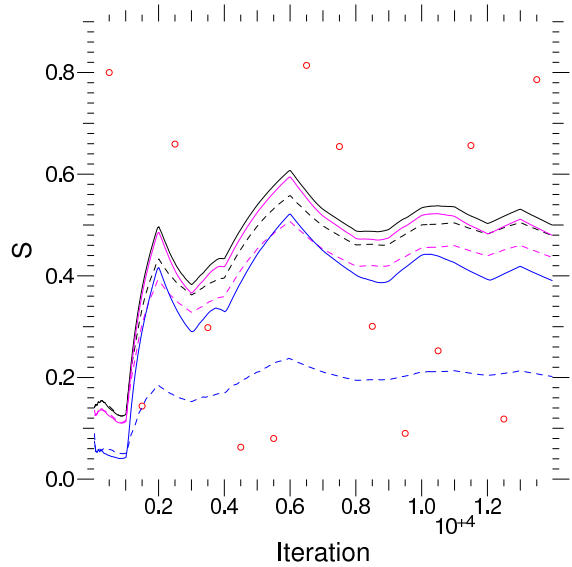}
\caption{
A 70-seconds AO sequence, during which the seeing, $\theta_0$, is modified every 1000$^{\rm th}$ iteration between 0.1'' and 1.8''. {\it Full lines\/}: Average Strehl when the defocus is adjusted during the sequence, {\it dotted lines\/}: the defocus is determined prior to the AO loop when $\theta_0=1.6''$. 
The magnitude of the reference star varies between 5 ({\it black\/}), 9 ({\it magenta\/}) and 11 ({\it blue\/}).  {\it Circles\/}: $\theta_0/2$ (Further parameter values: see right column of Table\,\ref{tab:simpar}). See the online edition of  JOSA A for the color version of this figure.}
\label{fig:noise}
\end{center}
\end{figure}

Both simulations represent extreme cases, since the initial defocus lengths are obtained under excellent seeing conditions and are thus exceptionally small.   
The black curves of Fig.\,\ref{fig:noise} show that  a defocus adjustment loop is not necessary for bright reference objects if one choses a constant and large defocus distance: the AO correction is then approximately the same for a constant and an adjusted defocus distance. 
Simulations and experimental results by O. Lai have likewise shown that in a low-order curvature system such as PUEO the AO loop performance is not very sensitive to changing defocus distances\,\cite{Lai}.    
However, for faint reference stars it is not satisfactory anymore to keep the defocus distance at a constant large value (see Fig.\,\ref{fig:noise}): aliasing effects require the defocus distance to stay at its inferior limit, and an adjustment to the current seeing is required. This suggests that, as the number of correction elements of AO systems rise and each correction element has to work with less photons, defocus adjustment loops might become increasingly necessary.

\section{Conclusion}

We have tested two different methods to automatically adjust the defocus length of a curvature sensor to the continuously changing turbulence conditions. 

In the first method the length is adjusted prior to the AO loop through the acquisition of a long exposure image and is then kept constant. However in extended observations of  astronomical objects or in any observation of fast moving targets, such as satellites, the turbulence intensity changes over the course of the AO sequence. For such observations, we have suggested a method to adjust the defocus length during the AO loop by use of the voltage values sent to the deformable mirror.

While the implications for AO in astronomy are apparent, it is also worthwhile in conclusion to note  the potential interest of this work for ophthalmology. Detailed information on the retina requires observations from outside the eye. Without optical correction the quality of such observations is markedly degraded by aberrations due to the various dioptric surfaces, such as the cornea, and by the aqueous and the vitreous humour. The correction needs to be dynamic, because of the eye's motions and because the aberrations fluctuate. Hofer et al.\,\cite{Hofer2001} suggest a required AO correction speed of 10-40\,Hz. 

Current methods to sense ocular aberrations, such as laser-ray tracing and  spatially resolved refractometry, are of limited applicability because they do not permit rapid acquisitions\,\cite{Moreno2001}. Dynamic aberration-correction is, therefore, only performed by use of Shack-Hartmann (S-H) sensors. Liang et al.\cite{Liang1997} have shown that S-H sensors permit fast, precise and repeatable measurements of ocular aberrations. On the other hand,  Moreno-Bruiso et al.\,\cite{Moreno2000} and Glanc et al.\,\cite{Glanc2004} found eyes where the S-H sub-aperture images are too distorted to be properly analyzed, and one needs to note that such eyes tend to be the most interesting from a clinical point of view. Porter et al.\,\cite{Porter2001} characterized the distribution of the ocular aberration in a population of 109 subjects and they concluded, in line with previous studies, that the ocular aberrations differed greatly from person to person - even though their study did not include subjects with pathologies such as cataracts or keratoconus. This suggests that retinal imaging could benefit substantially from the use of AO systems in which the sensitivity can be easily adjusted to the eye's aberration and their temporal fluctuations.

Curvature sensors have, in exploratory investigations, been successfully utilized to measure ocular aberrations \cite{Diaz2006}, \cite{Torti2008}. In these experiments, however, the sensors have not been used in combination with deformable mirrors, but were instead  derived in terms of a complex wavefront reconstruction algorithm. Since curvature-based AO systems in astronomy have proven simple and efficient when used together with a bi-morphic  or a membrane mirror\,\cite{Roddier1}, \cite{MACAO}, \cite{PUEO}, \cite{NICI}, \cite{Subaru}, we suggest to implement a curvature based AO system for retinal imaging. Compared to a S-H based system, the chief advantage of such a system will be the capacity to rapidly adjust the defocus distance when different patients, with their particular ocular aberrations are examined.

\section*{Acknowledgments}
We are happy to thank Jerome Paufique for helpful discussions, and the anonymous referee for his constructive corrections and suggestions. 
Financial support from the United States Air Force Office of Scientific Research is likewise acknowledged.

\begin{table}[htdp]
\begin{center}
\begin{tabular}{ l l l}  \hline \hline
Parameters & Sec.\,\ref{sec:constant} &  Sec.\,\ref{sec:var} (if different from Sec.\,\ref{sec:constant}) \\ \hline
& & \\
Telescope diameter, $D$ & 7.9\,m ; 220\,pixels & \\ 
Diameter of central obstruction & 1.0\,m ; 28\,pixels & \\ 
Size of influence functions, $M\times M$ & $237\times237$\,pixels & \\ 
Number of sub-apertures, $N$ & 80 & \\ 
& & \\
Size of phase screen & $2048 \times 256$\,pixels & $4096 \times 256$\,pixels\\ 
Seeing at 0.5\,$\mu$m, $\theta_0(0.5\,\mu$m)& [0.2'', 1.4'']& [0.1'', 1.8'']\\ 
Layer speed, $V$ & 2\,m/s & \\
Height of turbulent layer, $H$ & 100\,m & \\ 
Outer scale of turbulence, $L_0$ & $+\infty$ & \\
& & \\
Target magnitude, $m$ & 5 & \\ 
Sky magnitude, $m_S$ & 20 & \\ 
Photon flux at magnitude 0 & $10^{13}$\,m$^{-2}$\,s$^{-1}$ & \\ 
Atmospheric + instrumental throughput & 0.25 & \\ 
Imaging wavelength, $\lambda_I$ & 1.65\,$\mu$m & \\ 
Sensing wavelength, $\lambda_S$ & 0.7\,$\mu$m & \\ 
& & \\
AO loop frequency, $f$ & 1000\,Hz & 200\,Hz\\ 
Read-out latency & 0\,s & \\ 
Read-out noise and dark current & 0 & \\ 
First ... iterations disregarded  & 50 & \\ 
Minimal curvature radius & 13\,m & \\ 
of the deformable mirror &&\\
Loop gain, $g$ & [0, 1] increment: 0.1& \\ 
Extra-focal distance in an $f/60$ beam, $l$ & [0.1, 1.8] increment: 0.1 & [0.1,3.0] increment: 0.1\\ \hline

\end{tabular}
\caption{Parameter values of the numerical simulations, if not otherwise specified in the text.}
\label{tab:simpar}
\end{center}
 \end{table}

\end{document}